**ARTICLE**     Open Access

# Laser-written reconfigurable photonic integrated circuit directly coupled to a single-photon avalanche diode array

Giulio Gualandi[1,2], Simone Atzeni[2,3], Marco Gardina[2], Antonino Caime[1], Giacomo Corrielli[2], Ivan Labanca[4], Angelo Gulinatti[4], Ivan Rech[4], Roberto Osellame[2], Giulia Acconcia[4] and Francesco Ceccarelli[2 ✉]

**Abstract**
To date, most integrated quantum photonics experiments rely on single-photon detectors operating at cryogenic temperatures coupled to photonic integrated circuits (PICs) through single-mode optical fibers. This approach presents significant challenges due to the detection complexity, as cryogenic conditions hinder the development of scalable systems. In addition, going towards fully-integrated devices or, at least, removing the optical fibers would be also advantageous to develop compact and cost-efficient solutions featuring a high number of optical modes. This work reports on the direct coupling of a PIC, fabricated by femtosecond laser writing (FLW), and a silicon single-photon avalanche diode (SPAD) array, fabricated in a custom planar technology and compatible with the operation at room temperature. The effectiveness of this solution is shown by achieving perfect coupling and a system detection efficiency as high as 41.0% at a wavelength of 561 nm, which is the highest value reported to date among both heterogeneous/hybrid integrated and directly coupled systems. We also show the robustness of the coupling to misalignments, demonstrating that costly alignment procedures are not needed. Finally, we exploit the SPAD array to characterize a reconfigurable Mach-Zehnder interferometer, i.e., the basic building block of multimode reconfigurable PICs. This solution provides a new avenue to the design and implementation of quantum photonics experiments, especially effective when compact and cost-efficient systems are needed.

## Introduction

Quantum photonics experiments generally rely on three basic functions. The first is the generation of the quantum state, the second involves its manipulation, often through interferometric photonic integrated circuits (PICs), and the last stage involves the measurement or detection. Generally, in the field of discrete-variable quantum photonics, the detection stage is exploited for performing non-Gaussian operations on the quantum state by means of single-photon detectors. In particular, most experiments to date rely on the optical coupling of a PIC to either off-chip[1] and on-chip[2] superconducting nanowire single-photon detectors (SNSPDs)[3], which can provide exquisite photon detection efficiency (PDE), excellent dark count rate (DCR), and zero afterpulsing. However, cost, footprint, and power consumption of these detectors are often a limitation to the scalability of the experiment or even an obstacle to its implementation (e.g., when the detection is on a satellite). In addition, optical coupling is typically carried out through single-mode fiber arrays which introduce additional losses and whose geometry could be a limit to the arrangement of the waveguides at the circuit output[4].

Integrating single-photon detection into a PIC, working at or very close to room temperature, would play a major step forward in the scalability of the quantum hardware[5]. In this scenario, single-photon avalanche diodes (SPADs) are solid state detectors that can play a paramount role[6]. Experiments on the integration of SPADs and waveguide

Correspondence: Francesco Ceccarelli (francesco.ceccarelli@cnr.it)
[1]Dipartimento di Fisica, Politecnico di Milano, Milano, Italy
[2]Istituto di Fotonica e Nanotecnologie, Consiglio Nazionale delle Ricerche, Milano, Italy
Full list of author information is available at the end of the article circuits are blossoming both in the visible[7,8] and in the telecom range[9,10]. However, reaching a system detection efficiency (SDE), including the losses of both circuit and detector, higher than 10% with a fully-integrated solution still seems a daunting task. Nowadays, one of the most promising avenues towards a compact and scalable quantum photonic system featuring the highest performance consists in combining different integrated photonic technologies in a single heterogeneous/hybrid system[11]. Drawing inspiration from this approach, we report on the direct coupling of a four-mode PIC, fabricated by femtosecond laser writing (FLW) of waveguides in glass[12], to an array of thin silicon SPADs[6], manufactured in a planar custom technology[13]. The FLW technique[12] consists in the fabrication of waveguides by exploiting the permanent modification of the refractive index induced by a train of ultrafast laser pulses tightly focused in a transparent substrate. This technique appears to be particularly effective given its versatility and compatibility with three-dimensional geometries. FLW waveguides exhibit low propagation losses (around 0.1 dB/cm for both visible and infrared light), excellent coupling losses (even less than 0.3 dB), and a uniquely low birefringence (down to $10^{-6}$ [14]). Programmability is enabled by combining thermo-optic phase shifters (TOPSs) and thermal isolation trenches to reduce power dissipation and phase crosstalk, as required in high-fidelity universal photonic processors (UPPs)[15]. Regarding the detector, the custom technology developed at Politecnico di Milano for SPAD fabrication[13] has been proven effective to develop rugged thin SPADs featuring low power dissipation (thanks to a breakdown voltage lower than 40 V), combined with good detection efficiency (about 50% at 550 nm) and low dark counts (typically around kcount/s) when operating the detector at room temperature.

By fabricating each device with its own optimized custom technology and coupling them directly, we can get rid of bulky standard optical interconnections with no penalty in terms of losses. Indeed, we demonstrate a unitary coupling efficiency and a SDE as high as 41.0% for single photons at 561 nm wavelength, which is the highest value reported to date for both directly coupled and heterogeneous/hybrid integrated PIC-SPAD systems. In addition, we study also the resilience of this system, showing high tolerance to misalignment in any spatial direction. Finally, we report on the single-photon characterization of a reconfigurable Mach-Zehnder interferometer (MZI), demonstrating the feasibility of performing quantum photonics experiments by coupling a reconfigurable multimode PIC, like a UPP, with a SPAD array. Both the platforms we employ currently provide all the technological degrees of freedom needed to design a system optimized up to the ultimate limit, as required by the most demanding experiments in quantum photonics, and to provide compact and cost-effective solutions based on two-dimensional SPAD arrays and three-dimensional reconfigurable processors.

## Results

In the following, we briefly discuss the devices we employed for the experiment and then we present three different measurements that demonstrate the effectiveness of our solution.

The PIC (Fig. 1a) is composed of four planar waveguides, while the detector is an 8 × 8 SPAD array where four pixels were selected for this work (see Materials and methods). The PIC features a pitch of 127 μm at the input, to allow the coupling with a standard fiber array, and of 250 μm at the output, to have the same spacing of the SPAD pixels. The two external waveguides are independent and are intended for measuring the PDE of single SPAD pixels and to optimize the global alignment of the circuit to the SPAD array. On the contrary, the two central waveguides are part of a reconfigurable MZI, which consists of two nominally balanced directional couplers interconnected by straight waveguides (i.e., the two central arms), each of them having a dedicated TOPS. The MZI is depicted in Fig. 1a, where the entire layout of the device is visible. Since the electrical connection of the SPAD array is made by wire bonding, the output facet of the PIC was tapered as sketched in the artwork of Fig. 1b. By reducing the PIC front side from 12 to 1.3 mm it is possible to avoid any damage of the electrical bonding wires. An active alignment procedure (see Materials and methods) allows the simultaneous coupling of all four optical modes with the corresponding SPAD pixels. First, we perform a linearity measurement to accurately quantify the PDE of the SPADs at 561 nm wavelength, showing that it is possible to achieve a unitary coupling efficiency between PIC and SPAD array. Secondly, we perform a spatial measurement of the PDE as a function of the position of the waveguide on the SPAD active area and as a function of the distance between PIC and SPAD, demonstrating the robustness and tolerance of the coupling. Finally, we complete the characterization by showing the operation of a reconfigurable MZI coupled to two SPADs. All the measurements reported hereafter were performed by operating each SPAD pixel with an overvoltage of 5 V, unless explicitly stated otherwise.

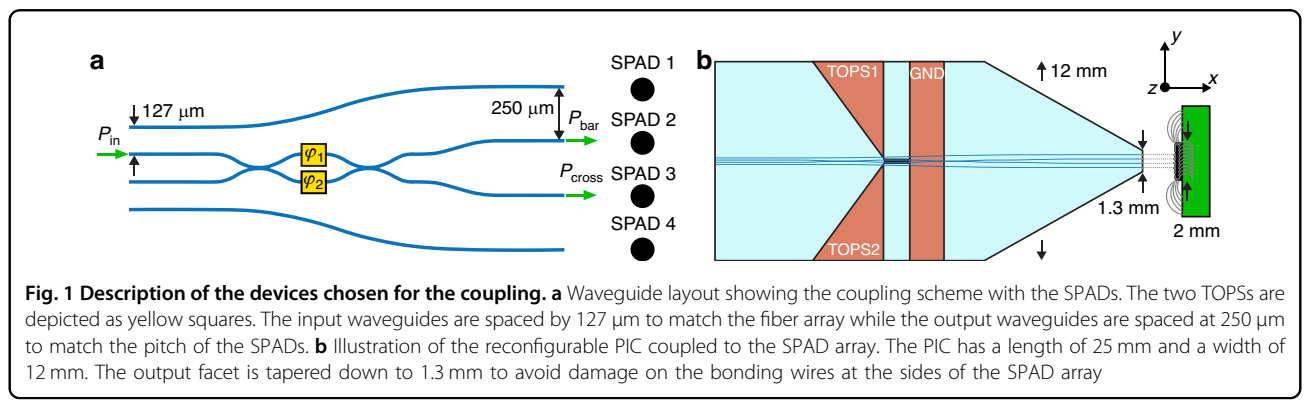

**Fig. 1 Description of the devices chosen for the coupling. a** Waveguide layout showing the coupling scheme with the SPADs. The two TOPSs are depicted as yellow squares. The input waveguides are spaced by 127 μm to match the fiber array while the output waveguides are spaced at 250 μm to match the pitch of the SPADs. **b** Illustration of the reconfigurable PIC coupled to the SPAD array. The PIC has a length of 25 mm and a width of 12 mm. The output facet is tapered down to 1.3 mm to avoid damage on the bonding wires at the sides of the SPAD array

### Linearity

The first step is a linearity measurement carried out on the independent waveguides at the sides of the MZI, corresponding to SPAD 1 and 4. Linearity measurements allow one to retrieve the PDE of a single-photon detector by measuring the relation between the photon rate at the output of the PIC, which was swept across 5 orders of magnitude, and the corresponding count rate measured by the SPAD. Therefore, such a measurement includes also the coupling losses, if present. Figure 2a, b depict the different count rates of SPAD 1 and 4, from which, through the linear fit (orange line), it is possible to infer the PDE of both, i.e., 45.3% and 42.4% for SPAD 1 and 4, respectively. The baseline value can be ascribed to the dark counts, which change from pixel to pixel, as discussed further in the "Materials and Methods" section. These measurements were also compared to the preliminary characterization performed on the SPAD array through a standard free-space setup[16], and in that case the PDE values were 40.5% and 41.1%, respectively. However, this discrepancy can be fully explained by the reasonably small disuniformity of the PDE on the active area of the SPADs, as highlighted in the next subsection. The disuniformity is also at the origin of the lower PDE that SPAD 4 features with respect to SPAD 1.

Next, we completed this step by measuring the linearity as a function of the overvoltage. Figure 2c reports the PDE of SPAD 1 for three different overvoltages, namely 5, 6, and 7 V, compared to the corresponding values measured with the free-space setup. As expected, the PDE increases with the overvoltage, reaching a maximum of 50.2%. The price paid for operating the SPAD at an increased overvoltage is a higher DCR, which moves from 16.6 to 25.4 kcount/s. Similarly, also the PDE of SPAD 4 increases with the overvoltage, reaching 49.6% at 7 V, while the DCR moves from 2.9 to 4.5 kcount/s. By including also the PIC losses (see Materials and methods), it is possible to calculate the SDE of the PIC-SPAD assembly, which increases from 37.0% to 41.0% for SPAD 1 and from 34.6% to 40.5% for SPAD 4, respectively, moving from 5 V to 7 V.

### PDE as a function of the PIC displacement

The second measurement aims at quantifying the robustness of the alignment procedure. Focusing our attention again on the two outermost waveguides, we started by characterizing the PDE as a function of the displacement of the PIC output facet with respect to the SPAD array. Figure 3a, b (and the corresponding plots for $z = 0$ reported in Fig. 3c, d) were acquired simultaneously as discussed in Materials and methods. Each picture shows the PDE on the surface of the SPAD, where the red circle delimits the active area (25 μm diameter). As clearly visible in the figure, the PDE at the edge of the active area

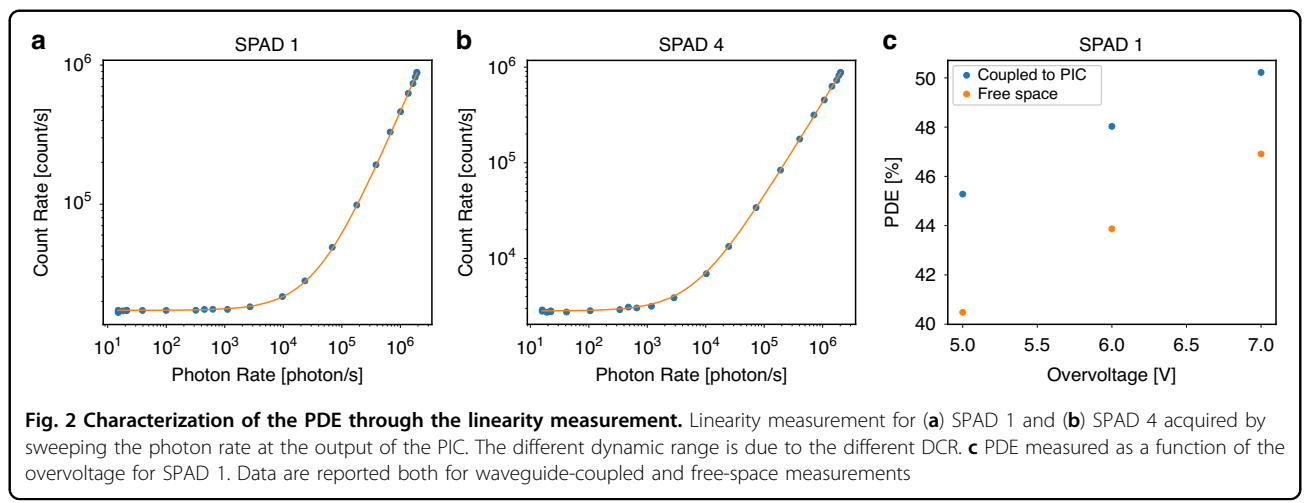

**Fig. 2 Characterization of the PDE through the linearity measurement.** Linearity measurement for (**a**) SPAD 1 and (**b**) SPAD 4 acquired by sweeping the photon rate at the output of the PIC. The different dynamic range is due to the different DCR. **c** PDE measured as a function of the overvoltage for SPAD 1. Data are reported both for waveguide-coupled and free-space measurements

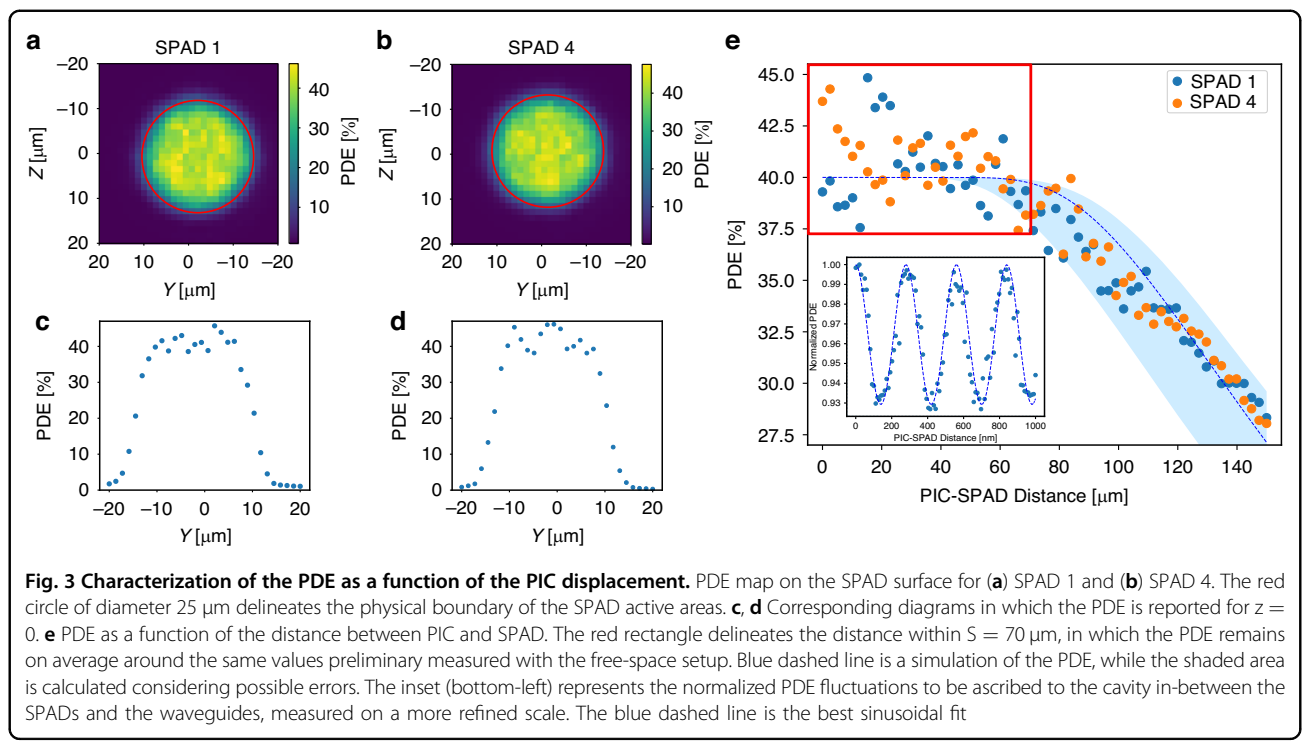

**Fig. 3 Characterization of the PDE as a function of the PIC displacement.** PDE map on the SPAD surface for (**a**) SPAD 1 and (**b**) SPAD 4. The red circle of diameter 25 μm delineates the physical boundary of the SPAD active areas. **c**, **d** Corresponding diagrams in which the PDE is reported for z = 0. **e** PDE as a function of the distance between PIC and SPAD. The red rectangle delineates the distance within S = 70 μm, in which the PDE remains on average around the same values preliminary measured with the free-space setup. Blue dashed line is a simulation of the PDE, while the shaded area is calculated considering possible errors. The inset (bottom-left) represents the normalized PDE fluctuations to be ascribed to the cavity in-between the SPADs and the waveguides, measured on a more refined scale. The blue dashed line is the best sinusoidal fit

is lower than in the center, in accordance with the lower electric field at the SPAD's boundaries, which results in a lower triggering efficiency in this region of the detector. Furthermore, it must be considered that the scan was performed with an irradiation spot of the size of the waveguide and therefore an area $1/e^2$ of $3.1 \times 3.4$ μm$^2$, so moving across the boundary there is a further decrease in PDE due to a portion of light falling outside the active area. However, for a large displacement of almost $d = \pm 10$ μm from the center, the PDE does not drop. This result shows how tolerant the coupling is to misalignments. Additionally, it is worth noting that, within the active area of the SPAD, the PDE is not uniform, reaching peaks of 46.1% and 47.6% for SPAD 1 and 4, respectively. Averaging the PDE values within the active area up to a diameter $D_{av} = 2|d| = 20$ μm yields average PDE values of 39.9 ± 3.3% and 40.4 ± 3.6% for SPAD 1 and 4, respectively. These PDE values are in good agreement with the ones measured without the PIC coupling. The small difference between the average value and the free-space characterization is well within the uncertainty of the different measurements.

The second experiment we performed was carried out to verify the dependence of the PDE on the distance of the PIC from the SPAD. For this measurement, we positioned the waveguides in the center of the SPADs at a negligible distance (in the order of 1 μm) from the surface. We then acquired the PDE values for a distance varying from 1 to 150 μm and reported the result in Fig. 3e. The values in the first S = 70 μm fluctuate from 45.0% to 37.5%. This is reasonable considering the nonuniformity of the PDE. Instead, the average values of the PDE within S are 40.3 ± 1.8% and 40.7 ± 1.5% for SPAD 1 and 4, respectively, thus in agreement with the previous measurement. As the distance between waveguide and SPAD is increased, the spot size of the impinging beam gets wider. However, up to about S there is no noticeable decreasing trend, thus demonstrating that also in this direction the coupling shows a remarkable tolerance to misalignments. Instead, the PDE begins to decrease beyond S, where it is reasonable to assume that the waveguide light spot has become wider than the area of the SPAD, and therefore some of the light does not impinge on the active area anymore. Figure 3e reports also a simulation of the PDE assuming photons diverging with the same distribution of an elliptical Gaussian beam having a waist of the same dimensions of the waveguide optical mode. Shaded lines are calculated considering possible errors in positioning the waveguides at the center of the SPADs and in the measurement of the waveguide mode (see Materials and methods). Simulations are in good agreement with the experimental data. These data allow us to infer also the numerical aperture for our waveguide NA ~ d/S = 0.14, which is reasonable for the low refractive index contrast typically induced by the FLW process[17]. For the sake of completeness, we performed also a measurement of the PDE as a function of the distance between the PIC and the SPAD by using a finer step size of 10 nm. This was done to characterize interference phenomena caused by the presence of the air cavity between the glass facet of the PIC and the SPAD surface, which could potentially contribute to the oscillations in the PDE observed in our results. The measurements are shown in the inset at the bottom left of Fig. 3e. As can be seen, the PDE exhibits sinusoidal interference fringes as the distance between the PIC and the SPAD is varied. A best-fit analysis reveals a fringe period of 280.5 nm, which is consistent with the expected half-wavelength of 561 nm for the light used in our experiment. The amplitude of the fringes is approximately 0.07, corresponding to a variation in PDE of about 3.2%. This demonstrates that the imperfect coupling due to the air cavity introduces a minor, yet measurable, contribution to the observed oscillations in the PDE. The same results were reproduced on four different positions on SPAD 1 active area.

Last, we have also completed the analysis by evaluating the effect of a rotation of the SPAD array around the Y-axis (see Fig. 1b). The results showed remarkable insensitivity also for this degree of freedom, with the PDE decreasing less than 10% for rotation angles below 3°. Given the radial symmetry of the SPAD active area, equivalent results hold also for rotations around the Z-axis.

### Operation of the reconfigurable MZI

The performance of the MZI is shown in Fig. 4a where the transmission of the two output waveguides of the interferometer is reported as measured by SPAD 2 (bar) and 3 (cross) as a function of the power dissipated by one of the two TOPSs. The experimental data follow the sinusoidal trend expected by Eqs. (2) and (3) (see Materials and methods). From the fit we obtained an initial phase $\varphi_0 = 5.06$ rad and a tuning coefficient $\alpha = 241.65$ rad/W, corresponding to a power required to reach $2\pi$, i.e., the full reconfiguration of the MZI, of 26.0 mW. However, to extract these values, we had to include in the model also background contributions $B_{bar}$, $B_{cross}$, and a nonunitary value for the amplitude $A$, which were found to be equal to 0.0595, 0.0374, and 0.903, respectively. The background values denote how the MZI cannot completely deplete the light from its arms, most likely due to an imperfect optimization of the directional couplers. However, these defects are not related to our assembly and can be easily addressed in FLW through preliminary fabrication runs dedicated to this wavelength[14]. As demonstrated by Fig. 4b, which reports the same data acquired by the SPAD detectors (orange dots) and the data acquired with standard power meters in linear regime (blue dots), the two measurements on the cross output are compatible in terms of background ($B_{cross} = 0.0272$), amplitude ($A = 0.916$), initial phase ($\varphi_0 = 4.98$ rad) and tuning coefficient ($\alpha = 241.76$ rad/mW). The same is true for the bar.

Finally, we wanted also to confirm that the coupling is not introducing crosstalk contributions external to the SPAD array, due as an example to photons exiting a waveguide and reaching an adjacent SPAD. To this aim, we measured the count rate of SPAD 1 (i.e., the one coupled to the adjacent independent waveguide) while we performed a phase sweep in the MZI. The count rate on SPAD 1 was measured simultaneously to the one on SPAD 2, and both the measurements are reported in Fig. 4c. While in SPAD 2 we can easily measure the sinusoidal curve, for SPAD 1, no change in count rate is appreciable on this scale. However, by looking at the count rate in a range between 16.6 to 18.0 kcount/s (Fig. 4d), it is possible to spot a modulation: when in the MZI waveguide there is a minimum, the count rate of SPAD 1 remains close to the DCR, while as soon as there is optical power transfer in the bar waveguide, the same sinusoidal trend is observed also on SPAD 1. The amplitude of this modulation, which is 0.31% (i.e., −25 dB) of the original one, is compatible with the internal optical crosstalk that is typically measured on SPAD arrays of the same generation[13].

### Discussion

In this paper, we presented the direct coupling of a PIC with an array of SPADs. Both devices, PIC and SPAD array, were manufactured by using custom technologies that make available all the degrees of freedom necessary to optimize the performance in each photonic task (i.e., manipulation and detection). Even more important, the optimization of each device can proceed independently of the coupling procedure, since the latter does not add technological constraints to the design of the two chips. The perpendicular arrangement chosen for the coupling is not even a limit for the scalability. The electrical connections of SPAD arrays can be managed effectively by shaping the PIC substrate to facilitate wire bonding, allowing one to scale up to several tens of pixels[16]. For larger arrays, a well-established solution involves through-silicon vias (TSVs), which connect each detector to the quenching electronics on a stacked configuration, eliminating wire bonding issues and enabling arrays with thousands of pixels[18]. While not currently implemented, the custom process used in this work can be adapted to include TSVs in the future. Regarding the PIC, electrical connections can be relatively simple, as wire bonding can be confined to the lateral sides. This design has already enabled large-scale UPPs up to 24 modes and 600 TOPSs[19]. However, as the number of TOPSs scales to thousands, routing all connections laterally may become impractical, necessitating a transition to flip-chip bonding[20]. Even the thermal management remains unaffected by the perpendicular mounting. Both SPAD arrays and PICs can use Peltier coolers to efficiently dissipate heat,

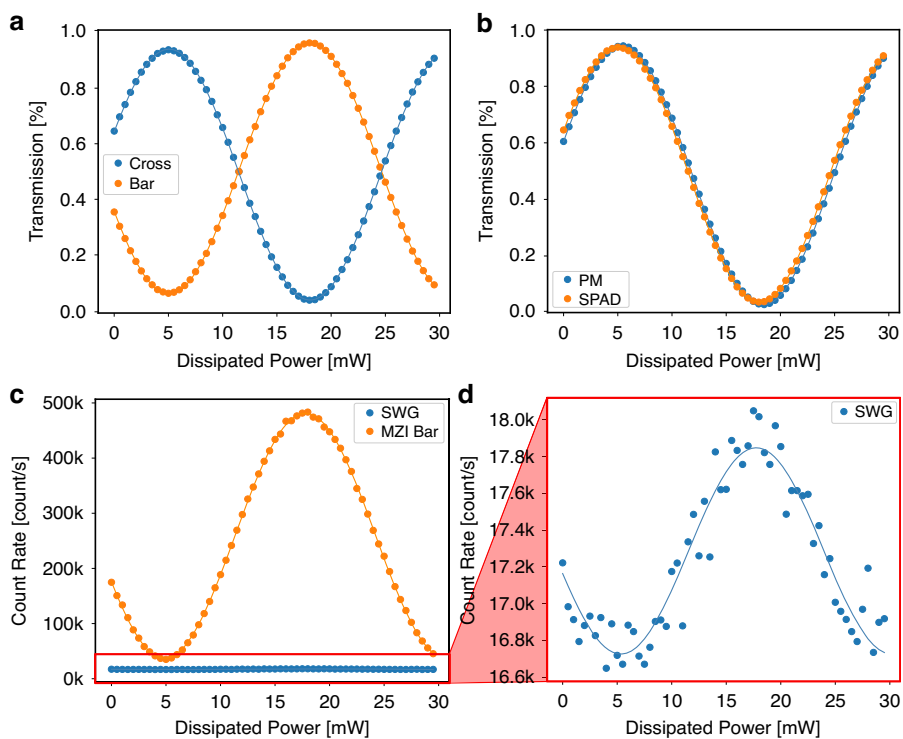

**Fig. 4 Characterization of the reconfigurable MZI. a** Normalized count rate measured on SPAD 2 (bar, orange dots) and SPAD 3 (cross, blue dots) as a function of the electrical power dissipated by the TOPS. **b** Comparison between the measurement performed with the SPAD (orange dots) and with the power meter (blue dots) on the cross output of the MZI. **c** Optical crosstalk measured by sweeping the power dissipated by the TOPS and by looking at the count rates of SPAD 2 (bar output of the MZI, orange dots) and SPAD 1 (independent straight waveguide carrying no light, blue dots) **d** Zoom in on the plot for a count rate between 16.6 and 18.0 kcount/s, where the optical crosstalk between the two SPADs is appreciable

ensuring compatibility with the electrical connections and maintaining optimal operating conditions for all components.

Moreover, it is worth also noting that the coupling technique shown in this work goes well beyond the two photonic platforms chosen for this experiment. Indeed, the simplicity of this approach allows one to envision also other SPAD devices, like dense CMOS SPAD imagers[21], for those experiments in which the number of optical modes is the most important parameter, or red-enhanced SPADs[22], for those experiments in which either reducing the losses or moving to red wavelengths is the main issue. One could also envision PICs fabricated in planar technologies (e.g., silicon nitride[23]) coupled to SPAD detectors. However, the full potential of this technique would be exploited only by coupling two-dimensional SPAD arrays with three-dimensional circuits[12], that are manufacturable thanks to the nonlinear mechanism at the origin of the femtosecond laser waveguide inscription process.

The reported measurements show that it is possible to achieve unitary coupling efficiency and negligible additional crosstalk on the adjacent optical modes. With the platforms chosen for this experiment, we demonstrate a PDE as high as 50.2% at 561 nm for an overvoltage of 7 V, which corresponds to a SDE = 41.0% by considering also the PIC losses. To allow an easy comparison of these results with the state of the art, we have summarized in Table 1 the best heterogeneous/hybrid integrated and directly coupled solutions reported to date. Our SDE is the highest among integrated PIC-SPAD systems[7–10,24,25]. This value is even higher with respect to the best waveguide-integrated SNSPDs[26] operating in the visible range, demonstrating that coupled systems can provide a viable path to scale quantum experiments in the short run, with no compromise on the SDE. In addition, it is worth noting that our system operates in a region of the spectrum in which single-photon sources at room temperature are already a reality[27]. This last point is of particular importance. Indeed, the integration of such sources into laser-written photonic circuits is already under development and has been proposed in the literature for testing quantum protocols on a space satellite[28], an application that represents an ideal target for the technology we are developing. This approach would allow the integration of all the three key functionalities—generation, manipulation, and detection—in a complete quantum photonic system for information processing without requiring cryogenic cooling. This breakthrough would pave the way for even more compact and scalable quantum technologies with a wide range of applications, both terrestrial and satellite-based.

Table 1　Comparison with the state of the art including both heterogeneous/hybrid integrated and directly coupled solutions (RT = room temperature, SDE includes propagation/coupling losses of the PIC with the exception of ref.[8], which does not report these data)

| Ref. | Technology | Photonic circuit | Detector | Temperature (K) | Wavelength (nm) | SDE (%) | DCR (count/s) |
| --- | --- | --- | --- | --- | --- | --- | --- |
| [26] | Heterogeneous | SiN | SNSPD | 1.4 | 532/1550 | 22/60 | 25–350 |
| [10] | Hybrid | Si | InGaAs/InAlAs SPAD | 220 | 1550 | 0.02 | 860k |
| [24] | Heterogeneous | Si | Ge-on-Si SPAD | 78 | 1310 | 13 | 279k |
| [8] | Heterogeneous | SiN | Si SPAD | RT | 488/532 | 10/6 | 107k |
| [7] | Heterogeneous | SiON | Si SPAD | RT | 850 | 5 | 100 |
| [25] | Coupling | Polymer | Si SPAD | RT | 785 | 2 | 5–30 |
| This work | Coupling | Glass | Si SPAD | RT | 561 | 41 | 4.5k–25.4k |

Additionally, measurements of the PDE across the active area and of the PDE as a function of the distance show us that we have several micrometers of tolerance for optimal coupling, both in terms of lateral alignment and distance from the SPAD. In fact, since the diameter of the waveguide is about one eighth of the diameter of the SPAD (one seventh if we consider the actual region in which the PDE does not drop), we can easily couple the waveguides of the PIC to the SPAD array without losing efficiency even for misalignments of several micrometers. This demonstrates that expensive equipment with nanometric resolution/accuracy is not needed and, on the contrary, the system proposed in this work is compatible with simple, compact, and cost-effective micropositioners. On the other hand, one potential improvement concerns the air cavity. While it can be exploited to optimize the coupling with SPADs featuring anti-reflection coatings designed for photons entering from air (as in our case), its presence makes the coupling slightly sensitive to nanometric displacements. If such oscillations cannot be tolerated in specific applications, the issue can be addressed by using index-matching oil along with an anti-reflection coating optimized for photons entering from glass. In addition, it is worth noting that the devices exploited for the experiment are also compatible with a permanent bonding process performed through optically clear adhesives, thus allowing one to remove the alignment stages during the operation to further compactify the setup.

Finally, to complete the work, we successfully characterized a reconfigurable MZI coupled to the SPAD array. This is an important step since, in the future, we aim at taking advantage of this technique for assembling SPAD arrays on FLW reconfigurable processors both in their planar[15] and three-dimensional[4] versions. Larger circuits will also bring higher losses and, in turn, lower SDE. However, the FLW platform allows the implementation of these processors with only a limited price to pay in terms of losses, as an example less than 3 dB for a 6-mode UPP[15], corresponding to a reasonably high SDE = 25.1%.

## Materials and methods

In the following, we briefly discuss the fabrication methods and the experimental procedures adopted throughout this work.

### Design and fabrication of the PIC

The fabrication of the PIC involved principally two steps: the fabrication of the optical circuit and the realization of the TOPSs. Regarding the first part, the waveguide inscription was performed by using a Pharos laser system, provided by Light Conversion. The laser beam was directed inside a 20x Zeiss water-immersion objective with 0.5 numerical aperture and then focused inside a borosilicate Corning Eagle XG glass substrate (1 mm thick) at 35 µm from the bottom surface of the sample. The waveguides were fabricated by using pulses at a wavelength of 1030 nm with a repetition rate of 1 MHz and a pulse duration of 180 fs. The fabrication process was purposely optimized for guiding light with a single mode at a wavelength of 561 nm. The final waveguides were inscribed with a power of 220 mW, a writing speed of 20 mm/s, and 6 overlapping traces and, after the laser writing process, the sample underwent a thermal annealing[14]. With this process, waveguides showed propagation losses around 0.11 dB/cm and a mode dimension $1/e^2$ of $3.1 \times 3.4$ µm$^2$. Next, deep isolation trenches were fabricated at the sides of the central arms of the MZI through laser ablation. This step is important to obtain

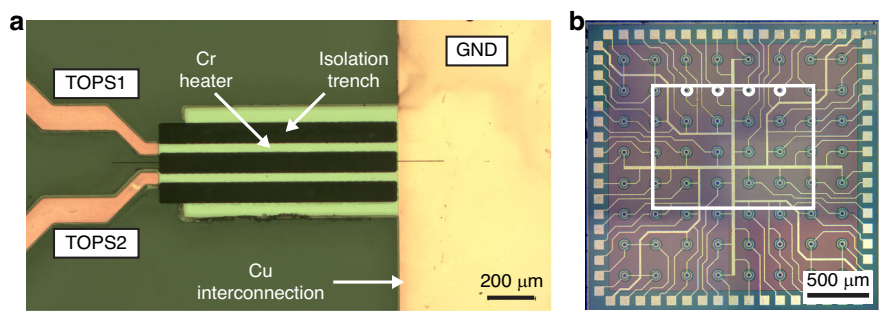

**Fig. 5 Microscope pictures of the two devices. a** Photomicrograph of the two TOPSs fabricated on the arms of the reconfigurable MZI. **b** Photomicrograph of the 8×8 SPAD array. The white rectangle represents the footprint of the PIC output facet (1.3 × 1 mm$^2$). White circles denote the SPADs (SPAD 1 to SPAD 4 from left to right) chosen for the coupling process

thermally-insulated waveguides and thus to reduce the power consumption of the reconfigurable MZI[29]. The trenches (Fig. 5a) are 1 mm long, 40 μm wide, and 300 μm deep.

Once the fabrication of the optical part was completed, the microheaters were fabricated by exploiting the process reported in ref.[30]. The fabrication consisted of a two-step photolithography process, in which chromium was the metal chosen for the resistive microheaters, while copper was the metal employed for interconnections and contact pads. We fabricated two microheaters, one on each MZI arm (Fig. 5a), showing electrical resistance of 440 Ω (top resistor in the figure) and 448 Ω (bottom resistor in the figure). Once the fabrication of the device was concluded, the substrate was cut by using a dicing saw to bring the facet down to a suitable dimension for safe contact on the SPAD array surface.

Finally, the PIC was mounted on an aluminum support along with two printed circuit boards (PCBs) hosting the electrical connectors. Aluminum bonding wires were employed to connect PCBs and PIC. Then, a fiber array was glued to the PIC input facet to obtain easy optical access and stable power at the input of the device. The transmission of the PIC after the fiber pigtailing process was as high as 81.6%, corresponding to 0.9 dB insertion losses.

### Design and fabrication of the SPAD array

The SPAD array (Fig. 5b) consists of 8 × 8 pixels, and it was fabricated exploiting a fully-custom planar technology[13] to optimize the performance. The fabrication was the result of a collaboration between Politecnico di Milano and the Institute for Microelectronics and Microsystems of the Italian National Council of Research (IMM-CNR sez. Bologna). The SPAD diameter is 25 μm, and the pixel pitch is 250 μm. The array features a common-cathode configuration allowing the same bias of all cathode terminals, while each anode can be independently connected to an external quenching circuit. In particular, in this work the four in-line pixels highlighted by the white box in Fig. 5b were selected to carry out the experiments (SPAD 1 to SPAD 4 from left to right), while all other SPAD anodes were conveniently biased to guarantee the off state of these pixels, thus avoiding any damage or crosstalk issue. The selected detectors feature a breakdown voltage of 34.1 V. The DCR can vary significantly from pixel to pixel due to the randomness of defect distribution across the wafer, as described for example in ref.[13]. For this type of array the DCR can range from a few 100's count/s for the best pixels, up to some 10's kcount/s for noisier SPADs. Each active SPAD was wire-bonded to one channel of an external fully-integrated active quenching circuit (AQC) array featuring the same structure reported in ref.[31]. The AQC array was fabricated exploiting a 180 nm CMOS technology by Austriamicrosystems (ams). The quenching circuit provides four independent output digital signals that are synchronous with the photon detection by the corresponding SPAD thus allowing single-photon counting on each channel. Both SPAD and AQC arrays were mounted on a host custom PCB (see next subsection) providing all the necessary bias voltages and extracting the digital outputs. The system was operated at room temperature.

### Experimental setup and data acquisition

The experimental setup, represented schematically in Fig. 6, was adapted from ref.[32]. A free-space 561 nm laser beam was fiber-coupled to an electronic variable optical attenuator (VOA, Thorlabs V450A), then the beam was split into two paths through a 3 dB fiber splitter (FS). One output was coupled to an auxiliary power meter (PM, Ophir Nova II with PD300R) acting as a power monitor, while the other beam was highly attenuated through a fixed fiber attenuator (FA) and again equally split into two fibers. In this way, we could inject light into two different input waveguides of the PIC at the same time.

First, we calibrated the number of photons exiting the PIC. Therefore, we measured the ratio between the power

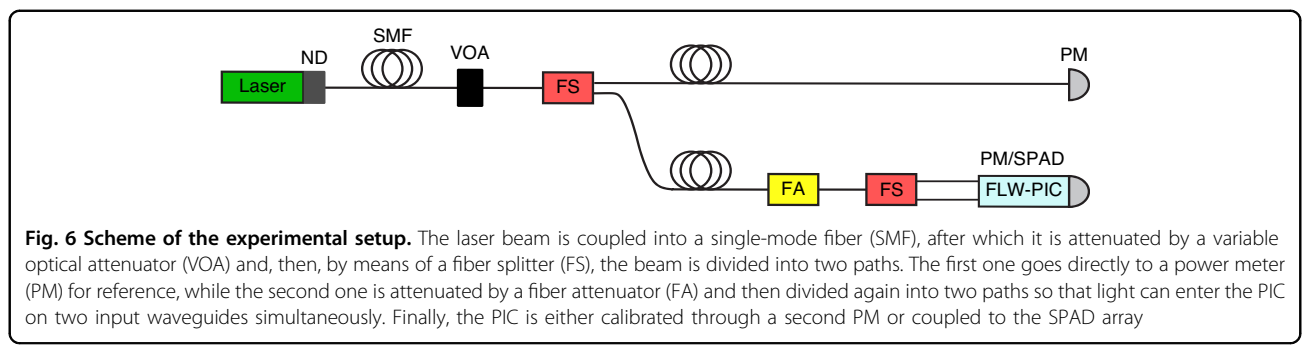

**Fig. 6 Scheme of the experimental setup.** The laser beam is coupled into a single-mode fiber (SMF), after which it is attenuated by a variable optical attenuator (VOA) and, then, by means of a fiber splitter (FS), the beam is divided into two paths. The first one goes directly to a power meter (PM) for reference, while the second one is attenuated by a fiber attenuator (FA) and then divided again into two paths so that light can enter the PIC on two input waveguides simultaneously. Finally, the PIC is either calibrated through a second PM or coupled to the SPAD array

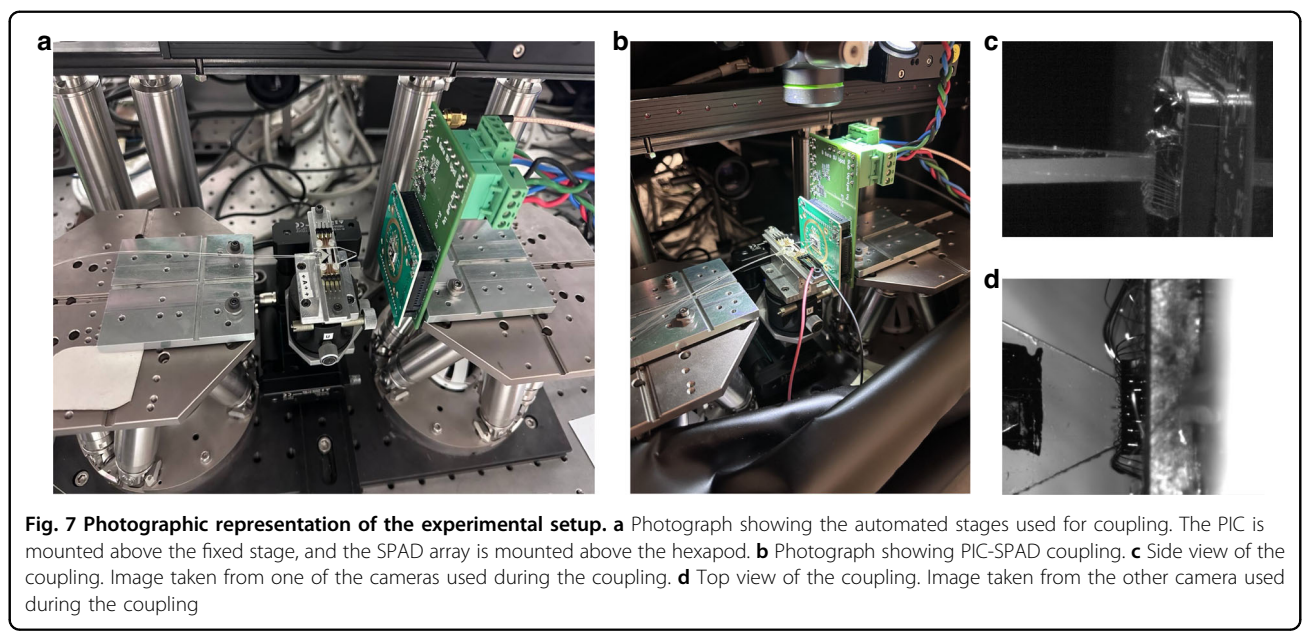

**Fig. 7 Photographic representation of the experimental setup.** **a** Photograph showing the automated stages used for coupling. The PIC is mounted above the fixed stage, and the SPAD array is mounted above the hexapod. **b** Photograph showing PIC-SPAD coupling. **c** Side view of the coupling. Image taken from one of the cameras used during the coupling. **d** Top view of the coupling. Image taken from the other camera used during the coupling

read by the auxiliary monitor and the power at the output of the PIC. To do so, the output beam was collimated with an aspherical lens (NA = 0.55) and steered to a second power meter. The measurement was carried out automatically by doing a sweep of optical power through the VOA and estimating the ratio through a linear regression. The resulting attenuation values were 60.0 dB and 59.7 dB for the output waveguides to be coupled to SPAD 1 and 4, respectively. In calculating the number of photons at the output of the PIC, we distilled the insertion losses of the lens, which turned out to have a transmission of 94.5%.

Then, power meter and collimation lens were replaced by the SPAD array. Two custom PCBs were designed: one on top hosted the SPAD array and the AQCs while the other one on the bottom was used to distribute the bias voltages, route the AQC output signals towards SMA connectors, and provide the necessary mechanical stability when fixed on top of a hexapod robot (Physik Instrumente). PIC-SPAD coupling was indeed performed thanks to these automated stages that provide six degrees of freedom, allowing for submicrometric alignment accuracy. A photon rate compatible with the dynamic range of the SPADs was guaranteed by introducing neutral density filters (ND) in front of the laser output, while allowing the auxiliary power monitor to read a signal with a signal-to-noise ratio reasonably high thanks to the fact the FA attenuated the light intensity only through the SPAD path (Fig. 6). The coupling was performed by optimizing the count rates measured by means of a two-channel frequency meter (Keysight 53220 A) connected to SPAD 1 and 4. Figure 7a, b show photographs of the whole experimental setup, while Fig. 7c, d show photographs of the devices taken by two cameras, equipped with magnification optics and positioned on the side and on the top of the setup, respectively. It is worth noting that this alignment procedure, based on the optimization of the count rate of the two lateral SPADs, results also in the alignment of the MZI on the corresponding central SPADs with no additional actions needed. Data acquisition was performed by recording the power read by the auxiliary monitor to estimate the number of photons exiting the PIC, and the photon count rate through the frequency meter operated in gated mode with gate time of 1 s, threshold of 1 V, and an input impedance of 50 Ω. Regarding the acquisition of the count rate data from the SPADs, we also considered the effect of the dead time, which was 55 ns. Therefore, to accurately estimate the PDE of the SPAD, we compensated for the dead time distortion as follows:

$$N_{phot} = \frac{N_{\text{det}}}{(1 - N_{\text{det}} T_{dead})} \quad (1)$$

where $N_{det}$ is the photon rate detected by the frequency meter, $T_{dead}$ is the dead time of the SPADs and $N_{phot}$ is the count rate reported in the Results section.

### Linearity

For the linearity, we performed an automated measurement on SPAD 1 and 4, where, through a Python program, we controlled the photon rate at the output of the two lateral waveguides through the VOA. After the acquisition of the count rates and the corresponding power measured by the monitor, we performed a linear regression on the data, assuming the intercept fixed to the measured DCR and estimating the PDE as the regression coefficient of the model.

### PDE as a function of the PIC displacement

The hexapod can be controlled by PC, thus enabling automated spatial measurements via a custom Python program. Regarding the spatial map of the PDE over the active area of SPAD 1 and 4, we took a total of 30 × 30 measurements, in a total area of 40 × 40 μm². The program was designed in such a way that the hexapod moves to the first-row value, i.e., 20 μm, reaching in 30 measurements a value of −20 μm (from left to right) and repeats this operation for all the 30 columns. In a similar fashion we also measured the PDE as a function of the distance, i.e., once positioned in the center and almost in contact with the SPADs (distance in the order of 1 μm), the hexapod was programmed to move away by performing 60 measurements over a total distance of 150 μm. During both measurements, the count rates of SPAD 1 and 4 and the power of the monitor were acquired simultaneously, along with the displacement coordinates of the hexapods. These measurements were performed by keeping the laser intensity fixed and thus the calculation of PDE was done through a single acquisition.

### Simulation of the PDE as a function of the PIC displacement

For the simulation of the PDE as a function of the distance between the detector and the waveguide facet, we considered a circular SPAD with a diameter of 25 μm. The active area of the detector was set to have a flat response equal to the mean PDE reported in the previous sections. Detector response was set to zero outside the active area. The beam at the output of the waveguide was approximated as an elliptical Gaussian beam with mode field diameters $w_x = 3.1$ μm and $w_y = 3.4$ μm. The intensity distribution was normalized so that its integral was always equal to unity. The system response was then calculated as the product between the normalized intensity profile of the Gaussian beam and the detector PDE for different values of the PIC-SPAD distance. Shaded lines are calculated considering the minimum and maximum values of the system response for an offset $\delta = \pm 3$ μm between the center of the SPAD and the center of the waveguide, as well as a possible error $\Delta w = \pm 0.3$ μm in the experimentally retrieved values of $w_x$ and $w_y$. The dashed blue line is calculated considering no error in the system (i.e., $\delta = 0$ and $\Delta w = 0$).

### Operation of the reconfigurable MZI

For characterizing the MZI response, we injected light into one of the two input waveguides (Fig. 1a) and measured the count rate of SPAD 2 and 3 as a function of the power dissipated by a microheater (the upper one, see Fig. 5a). The latter was connected to a source meter (Keysight B2902A) to perform a power sweep from 0 to 30 mW, with a step of 0.5 mW. The count rate values were measured simultaneously on SPAD 2 and 3 and then, they were normalized over their sum. For fitting the data, we used sinusoidal models with some modifications that we introduced to include the nonidealities of our MZI:

$$N_{bar} = B_{bar} + A \sin^2\left(\frac{\varphi_0 + \alpha P}{2}\right) \quad (2)$$

$$N_{cross} = B_{cross} + A \cos^2\left(\frac{\varphi_0 + \alpha P}{2}\right) \quad (3)$$

where $N_{bar}$ and $N_{cross}$ are the normalized count rate values for bar and cross SPADs, $B_{bar}$ and $B_{cross}$ are the corresponding background count rates, $A$ is the amplitude of the interference fringe, $\varphi_0$ is a constant initial phase due to fabrication tolerances, $\alpha$ is a linear tuning coefficient and $\mathcal{P}$ is the electrical power dissipated by the microheater. Note that the phase grows linearly with the dissipated power, in agreement with the fact that microheaters were fabricated with chromium, a material whose resistivity has negligible dependence on the temperature[30].

### Acknowledgements
G.A. and F.C. acknowledge financial support from the project HI-LIGHT (Hybrid Integration of Laser-written Interferometers and sinGle pHoton deTectors), grant n. 2022JRSST2, funded by the Italian Ministry of University and Research (MUR) through the PRIN 2022 program (D.D. n. 104, 02/02/2022). G.C., R.O. and F.C. acknowledge financial support from the project EPIQUE (European Photonic Quantum Computer—Grant Agreement No. 101135288) funded by the European Union's Horizon research and innovation program. The fabrication of the photonic integrated circuit was partially performed at PoliFAB, the micro- and nanofabrication facility of Politecnico di Milano (www.polifab.polimi.it). M.G., A.C., R.O. and F.C. would like to thank the PoliFAB staff for the valuable technical support.

### Author details
[1]Dipartimento di Fisica, Politecnico di Milano, Milano, Italy. [2]Istituto di Fotonica e Nanotecnologie, Consiglio Nazionale delle Ricerche, Milano, Italy. [3]Integrated Quantum Optics Group, Institute for Photonic Quantum Systems (PhoQS), Paderborn University, Paderborn, Germany. [4]Dipartimento di Elettronica, Informazione e Bioingegneria, Politecnico di Milano, Milano, Italy

### Author contributions
G.A. and F.C. conceived the original idea. S.A. and F.C. designed the experimental apparatus. G.G., S.A., G.C. and F.C. carried out the experiments and analyzed the data. G.G., M.G., A.C. and G.C. fabricated the PIC. I.L. and I.R. designed the electronics operating the SPAD array. A.G. performed the design and the preliminary characterization of the SPAD array. R.O., G.A. and F.C. supervised the project and provided the financial support. G.G. and F.C. wrote the original draft of the manuscript. All authors contributed to final version of the manuscript.

### Data availability
Data are available upon reasonable request.

### Conflict of interest
G.C., R.O. and F.C. are co-founders of Ephos. The other authors declare no competing interests.